\documentclass[twocolumn,twoside,slac]{revtex4}
\usepackage{graphicx}
\usepackage{fancyhdr}
\begin{document}

\title{A Model for Grid User Management}

%

\author{Richard Baker}
\affiliation{
        RHIC/USATLAS Computing Facility \\
        Department of Physics \\
        Brookhaven National Laboratory\\
        Upton, NY 11973, USA}
\author{Dantong Yu}
\affiliation{
        RHIC/USATLAS Computing Facility \\
        Department of Physics \\
        Brookhaven National Laboratory\\
        Upton, NY 11973, USA}
\author{Tomasz Wlodek}
\affiliation{
        RHIC/USATLAS Computing Facility \\
        Department of Physics \\
        Brookhaven National Laboratory\\
        Upton, NY 11973, USA}

\begin{abstract}
Registration and management of users in a large scale Grid computing environment presents
new challenges that are not well addressed by existing protocols.  Within a single Virtual
Organization (VO), thousands of users will potentially need access to hundreds of computing
sites, and the traditional model where users register for local accounts at each site will
present significant scaling problems.  However, computing sites must maintain control over
access to the site and site policies generally require individual local accounts for every
user.  We present here a model that allows users to register once with a VO and yet still
provides all of the computing sites the information they require with the required level of
trust.  We have developed tools to allow sites to automate the management of local accounts
and the mappings between Grid identities and local accounts.

\end{abstract}

\maketitle

\thispagestyle{fancy}

\section{\label{introduction-ref}Introduction}
Grid computing for the high energy and nuclear physics communities will
need to deal with thousands of users and hundreds of computing 
facilities in the near future. Traditional computing models require that 
each user must register individually with each computing facility.
This model will not scale as the numbers of users and facilities increase.
Some tools have already been developed to address the scalability problems
by centralizing user registration with a central service for a distributed
organization including CAS \cite{cas-ref} and 
VOMS \cite{voms-ref}. As they are currently implemented, these approaches 
do not provide enough information about users to allow access to some sites,
especially DOE laboratories. The separation of user management 
software from the user policy creates potential security problems 
for the sites.
The purpose of this project was to adapt the existing Grid mapfile tools 
to the large laboratory environment.  The particular 
problem that we 
intend to address is the need for strong pre-registration of users, and 
we believe that this builds nicely on some of the preliminary pieces that 
were available as early as 2001.
Our focus has been on developing a user account management system and tools 
to allow sites to keep track of users.  We intend all of these developments 
to be compatible with whatever VO management tools are adopted for 
the Large Hadron Collider Computing Grid Project (LCG), 
although we do have some requirements to put on those tools.

Before a user can be authorized to use resources at a site, the site 
must have some basis for trusting the user and approving the request. 
For a site to perform user based authorization, the user must 
be known to the site.  Different sites may have different requirements 
that must be satisfied before granting site access, but at a minimum,
some basic user information must be collected and provided to the site 
by secure, trusted and auditable means.  An additional requirement is 
to maintain history information of site access rites. To get started on 
solving these problems,
we developed a user management model
that has been presented at several open
meetings over the past year, most strongly
within the PPDG Site-AAA working group.

A primary requirement is that the user information is trustable, and this
requires that the information is validated before it is distributed to 
resource providers.  To accomplish this, a Virtual Organization must 
identify some set of members who will act as 
Registrars, meaning that they are the ones who can add users into the 
VO's membership list.  This list of registrars must be known and accepted 
by the resource providers who will be providing services to the VO.
When a user joins the VO, she/he goes to one of the designated registrars 
and provides basic identification and contact info such as Grid Certificate, 
Distinguish Name(DN), real name, institution and email address.  The exact info 
to be provided must be agreed among all of the resource providers.  The user 
also "signs" a User Agreement that is required by the VO.  LCG is beginning to 
work on this registration model, and a prototype User Agreement is being used.

The registrar then adds the user to a VO database.  The user record will
include the information that is required by the resource providers plus the name
of the registrar.  Existing VO management tools do not
keep track of enough user info to satisfy many resource providers, and 
better user registration and VO management tools will need to be developed.

Once the user is added to the VO database,
sites will need tools to download and keep
track of these users.  This requires more than
just a simple "make-gridmapfile" tool.  Sites
need to be able to plug in local policy modules
to control the user access.  Historical access
rights need to be kept track of.

\section{\label{requirement-ref}VO User Management Requirements}
Sites need to serve large sets of Grid users and 
many sites require pre-registration of Grid users. Grid  
users need access to a large number of sites without 
the detailed  knowledge about these individual sites. 
To make the whole transaction scale to the Grid environment,
sites and VOs should work out reliable Grid 
user registration mechanisms.  Grid user registration process 
includes three parties: virtual organization (VO), Grid users 
and resource providers.  Virtual organization needs to satisfy
the requirements of existing large Grid resource sites with 
respect to the acceptance of Grid credentials for access to their 
services.  Because those requirements are non-uniform,  
the user registration mechanisms and 
toolkits must provide to sites an interface to insert their 
own implementations of their requirements based on site policies.  
Sites must articulate 
what resources they are making available as Grid resources and the 
details of the access requirements.  This user registration mechanisms 
should enable sites to restrict access to the members of a specific VO,  
not automatically grant access to anyone with authentication credentials 
from Grid. 
The requirements for these three parties are:
\begin{itemize}
\item {\bf Site Requirements}:
All sites currently have authentication systems for identifying users 
and verifying user identities.  Any new user registration system must
be capable of being integrated with site policy and site infrastructure.
Sites need to collect sufficient information about 
users and the registration chain. Information must be provided to site in 
secure, trusted, auditable manner.  The Grid user list to sites should 
be reasonably static.
\item {\bf User Requirements}:
Users should be required to register only
once per virtual organization.  The registration process must be reasonably local
to the user. Virtual organizations should only collect reasonable information
without violating the user's privacy. 
The items collected by VO should be relatively 
static.  All of the private user information should be protected.
\item {\bf VO Requirements}
Sites must have reasonably complete and up-to-date user list so that 
the Grid scheduler can distribute user jobs without keeping track
of all users' authorization information at each individual site.
\end{itemize}
\section{\label{architecture-ref}Architecture of VO User Account Management Tool}
\begin{figure*}
\centerline{
 \scalebox{0.55}
 {\includegraphics {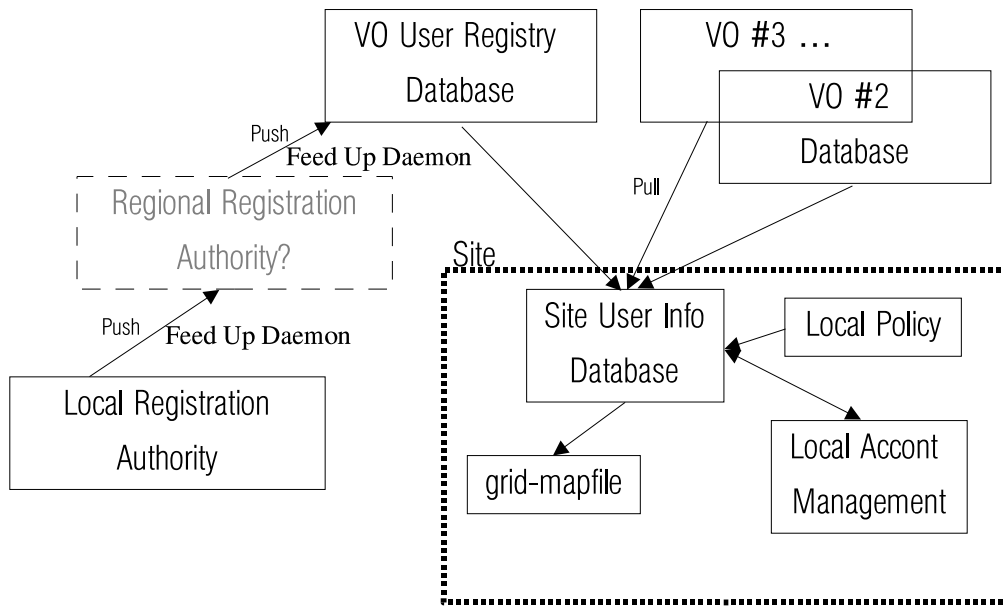}}}
\caption{Schematic Diagram of VO server and Grid User Management System}
 \label{figure:schematic}
\end{figure*}

The Grid user authentication infrastructure for virtual organization 
consists of two components: VO for large collaboration and Grid user 
management system for sites. The detailed implementation of  these 
two components  can be found in the remaining part of this section. 
Figure \ref{figure:schematic} shows
how these two components are organized together to enable large number 
of Grid users to register with distributed site resources. 
The user registration authority
obtains the user information and sends this information to upper-level
registration authority via a feed up daemon. Then the user information 
is stored at VO user database. The local site daemon interacts
with the VO user database to obtain user information and stores into 
the local database. The local policy is considered  on whether 
the user can be registered at the local resource provider. If the 
user gets authorization, the Grid user management system interacts
with the local user account management system on behalf of the user.
In this section, we also discuss the security issues with the infrastructure 
 and provide a solution based on Grid security infrastructure (GSI).

\begin{figure*}
\centerline{
 \scalebox{0.55}
 {\includegraphics {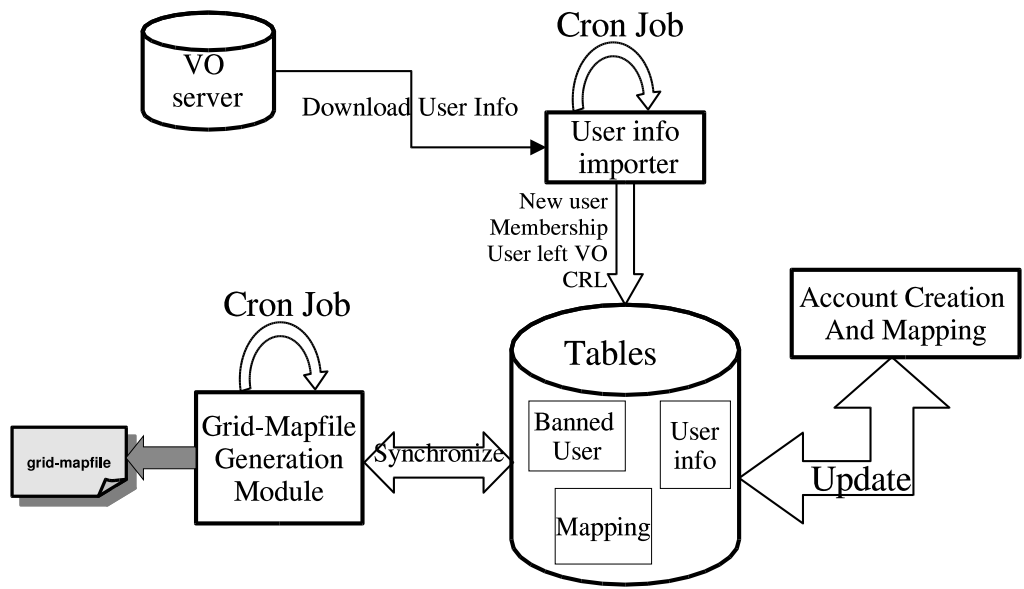}}}
\caption {Grid User Management System Architecture
 \label{figure:architecture}}
\end{figure*}

\subsection{Virtual Organization System}
{\bf VO User Database}: an expandable columnar database that 
contains one record per authorized user. Each record includes
sufficient information 
to allow account creation at remote sites. Exact content of record is 
subject to negotiation between sites and VOs. The VO user databases 
must  be secure both to safeguard potentially private user information
but also to ensure a reliable and trustable source of information for sites.
The VO user database will accept inputs from the remote user registration 
authorities in a secure and (two-way mutually authenticated) trusted 
way. Such inputs will be pushed from the registration authorities to 
the VO user database. The VO user database must also provide a capability
for assigning users to one or more groups within the VO and this group 
membership information must be available to the sites together with 
other user information.

{\bf VO User Registration Authority}: it basically tracks the same 
information about 
users as the primary VO user database and feeds this information 
up to the VO. The 
purpose is to distribute the responsibility of registering users. The 
functionality may be tiered with sub-registration authorities feeding
up a chain. Any user registration authority could also function as a VO
User database for a regional subsection of a VO, so the same implementation
could be used for both the root VO and regional sub-VO, with the feed 
up communications handled by an 
external daemon. A local User registration authority may contain more 
information about a user than is required by the VO user database, and 
the feed up daemon would only push the information that is actually required.

{\bf Feed Up Daemon}: This module establishes the secure, mutually-authenticated
connection and pushes the required user information into the VO user database.

\subsection{Grid User Management System for Sites}
Figure \ref{figure:architecture} provides the system
 architecture of local Grid user management system and how its components
 interact with each other to register the globally dispersed users to sites
 and allow them to use the site  resources.

{\bf Site Daemon}: This is the active piece that runs at each site to pull 
user information from a VO user database and invoke the local tools that track
and manage local accounts. 
This daemon must use secure mutually-authenticated
protocols to pull the user information from the VO server. The information 
required by the site may be a subset of the information available, and only 
the required information should be retrieved. The site daemon uses the 
information  retrieved from the (one or more) VO user database together with 
local information such as a list of banned users or a list of existing 
non-Grid users 
 to initiate and control the local functions described in the following
 paragraphs.

{\bf Local Database}: Maintain a database tracking user information, including all of the 
information required by the site to create user accounts 
plus any additional local customization that may be required. This database
 should serve both current and historic information.

{\bf Create and maintain local accounts}: This could mean different things at 
different sites, anywhere from directly invoking tools that actually 
create accounts to printing the user information for a human administrator
to review and maintain accounts manually. The diversity of possible implementations  requires a send request followed by an asynchronous process that 
obtains the results of the requests, possibly many hours later. The results
can be (at a minimum) success or failure of the request. The actual local
user identity assigned to the request is also returned as part of the results
if it is a new account request.

{\bf Maintain the grid-mapfile}: 
This means creating a record for each user with the user certificate.
The user's distinguished name is mapped to local account name, and the 
 mapping  information is written into a system file belonging to the 
Grid gatekeeper.
If a new account request has not been successfully completely yet, 
obviously no grid-mapfile record is created for that user yet. 

\subsection{Security Consideration}
The Grid user registration system satisfies the requirements listed in Section
\ref{requirement-ref}. When the sites fetch user information from a VO 
server, it uses the Grid security infrastructure (GSI) \cite{GSI} to mutually
authenticate with the VO server. The site administrator must register his Grid 
certificate at the VO server. After he/she is authenticated as a legal 
site administrator, his Grid identify will be enrolled in the  VO site 
administrator list. Then he is authorized to download user information. 
The LDAP-based VO \cite{ldap-1,ldap-2}
server can use access control list (ACL) with different granularities for 
different site administrators. On the other hand, even compromising the Grid
user management system could not grant illegal access to other resources.  
\section{\label{manual-ref}Software Tools for VO Management and Site Administration}
The name of our software is ``GUMS'', the acronym for the Grid User 
Management System. In this section, we will provide the install 
instructions and user manual for GUMS. Before you install GUMS, 
you should obtain the VO server information from the VO administrator 
of your collaboration. You also need to install and setup  
a MySQL server at your site. The MySQL server and the GUMS software 
 do not need to be installed on the same host. 
 
\subsection{Building/Installing GUMS}
The software package can be obtained without any restriction at 
 Brookhaven National Laboratory\footnote {http://www.atlasgrid.bnl.gov/testbed/gums/.}.
After you obtain the software, you only need to execute the 
following commands to install and configure the software.

{\bf Install from tarball}
\begin{itemize}
\item Obtaining and Unpacking the Package's source code
\begin{verbatim}
  $ tar zxvf  gums-1.2.src.tar.gz
  $ cd gums-1.2
\end{verbatim}
\item Configuration
\begin{verbatim}
  $ ./configure [--prefix <PREFIX>]
\end{verbatim}
\item Building
\begin{verbatim}
  $ make
\end{verbatim}
\item Installation
\begin{verbatim}
  $ make install
\end{verbatim}
\end{itemize}

{\bf Install from source RPM}
\begin{itemize}
\item        Obtaining and Unpacking the Source Package
\begin{verbatim}
  $ rpm -i gums-1.2-1.src.rpm 
\end{verbatim}

\item Building
\begin{verbatim}
  $ cd /usr/src/redhat/SPECS
  $ rpm -ba gums-1.2.spec
\end{verbatim}

\item Installation
\begin{verbatim}
  $ rpm -Uvh gums-1.2-1.i386.rpm  
\end{verbatim}
\end{itemize}

{\bf Post-Installation} \\
You need to setup your local MySQL database before you use 
the system commands.  To set up the configuration file,
go to your installation directory. If you install the GUMS rpm, 
the installation directory by default is
/usr/local/GUMS/, the configure file is /usr/local/GUMS/etc/VO.conf,
 and the executables are in /usr/local/GUMS/sbin/.
\begin{itemize}
\item  Select the VO LDAP server from which you could get user 
information and certificates. By defaults, it is
       ldap://spider.usatlas.bnl.gov:6200/ou=us-atlas,o=atlas,dc=ppdg-datagrid,dc=org. 
\item You need to choose a MySQL server to save your user information
       and certificates locally.  Before you run any command, make sure
       your MySQL is started and you have proper permission.
\item Initialize MySQL database, set up database, database tables and
        populate the table with records by running the command ``initdb".
\end{itemize}

\subsection{Command Line Tools}
The following tools are available for the GUMS software.
\begin{itemize}
\item {\bf initdb:} it initializes the local MySQL database. 
 This command setups the MySQL user account, creates tables, populates 
the tables with 100 local Grid accounts (grid001 to grid099). 
\item  {\bf getVOusers/secgetVOusers:} these commands download the 
user information from the VO servers specified by the configuration file, 
update the local Grid account databases.  If a new user comes 
from a VO, his/her information is inserted to the database, 
the information about his role in VO is also be put into the user 
database.  Based
on his role, the user obtains the site authorization
to execute his requests which abide by his role. 
At the current implementation, the
role authorization is implemented based on UNIX  group. For each
role, a group is created with specific authorization 
for that role. These two command line tools keep track of all members
of a VO server. If a member resigns from the VO, then the
tools will disable the corresponding authorization assigned
to this user by removing the user from the group. These
tools send email to the site administrator for adding and
removing a user to/from a group. The site administrator
should be involved to do the actual operations.

\item {\bf updategroup:} it checks the user's VO memberships 
and roles, and assigns the user to the UNIX group created for 
each role in the VO.  The group is used to implement role based 
authorizations in a VO.  For example, an ATLAS VO has three roles: 
simulation role, reconstruction role, and analysis role. We create 
three UNIX groups for these roles. Each group can use the computing 
resource and disk space according to their group allocations. 
If a VO member has the role for reconstruction, 
his/her account belongs to 
the UNIX group "reconstruction" and he is authorized to do work 
on the computing resource allocated for reconstruction.

If this tool runs in "cron" mode, it scans the pending group requests 
and sends the requests of updating the users' group membership to the 
site administrator.  If a new role in a VO is found, it sends 
emails to the site administrator to create a group for this new role. 
After the site administrator creates the group, he/she needs to insert 
the newly created group to the Grid user database. This tool provides  
the interface for the site administrator to access the database.

\item {\bf generate\_gridmap:} This script scans through all of user 
records stored in MySQL database and generates a gridmap file.
\end{itemize}


\section{\label{conclusion-ref}Conclusions and Future Work}
In this paper, we discussed requirements for distributed,
scalable Grid user registration.  We presented a system
we developed, called GUMS, which automates user registration
and management at for a Grid site.
GUMS is designed to satisfy the stated distributed
registration requirements while incorporating site policies.
Our experience using GUMS proved that this system was scalable, 
flexible and secure.   

The distributed registration infrastructure and the trust relationships
within a virtual organization are largely undeveloped at this time, but
organizations such as the LHC (The Large Hadron Collider) Computing 
Grid Project are starting to address this issue.
The current implementation of GUMS is based on plain password protected
MySQL, but MySQL recently included the 
Secure Sockets Layer (SSL) and  the x509 certificate in its 
authentication module. We will incorporate these security enhancements
in the MySQL-based local
database which stores user information for GUMS. 

\begin{acknowledgments}
The authors wish to thank RHIC/USATLAS computing facility group 
for their valuable comments and discusses for this work.
This work is supported by PPDG, PPDG Site AAA and ATLAS grants.
\end{acknowledgments}

\end{document}